\documentclass[aps,pra,twocolumn,longbibliography,showpacs,superscriptaddress]{revtex4-1}%
\usepackage{bm,amsmath,amssymb}
\usepackage{amsfonts}
\usepackage[english]{babel}
\usepackage[colorlinks=true,linkcolor=blue,citecolor=blue,urlcolor=blue]{hyperref}
\usepackage{xcolor}
\usepackage{enumitem}

\usepackage{txfonts}

\usepackage{graphicx}

\begin{document}
\title{Measuring geometric phases with a dynamical quantum Zeno effect in a Bose-Einstein condensate}

\begin{abstract}
A closed-trajectory evolution of a quantum state generally imprints a phase that contains both dynamical and geometrical contributions. While dynamical phases depend on the reference system, geometric phase factors are uniquely defined by the properties of the outlined trajectory. Here, we generate and measure geometric phases in a Bose-Einstein condensate of $^{87}$Rb using a combination of dynamical quantum Zeno effect and measurement-free evolution. We show that the dynamical quantum Zeno effect can inhibit the formation of a geometric phase without altering the dynamical phase. This can be used to extract the geometric Aharonov-Anandan phase from any closed-trajectory evolution without requiring knowledge or control of the Hamiltonian.
\end{abstract}
\author{H. V. Do}
\thanks{These authors contributed equally to this work.}
\address{European Laboratory for Non-Linear Spectroscopy (LENS),
University of Florence, via N. Carrara 1, 50019 Sesto F.no (FI), Italy}
\author{M. Gessner}
\thanks{These authors contributed equally to this work.}
\address{European Laboratory for Non-Linear Spectroscopy (LENS),
University of Florence, via N. Carrara 1, 50019 Sesto F.no (FI), Italy}
\address{QSTAR and CNR-INO, Largo Enrico Fermi 2, 50125 Firenze, Italy}
\address{D\'{e}partement de Physique, \'{E}cole Normale Sup\'{e}rieure, PSL Universit\'{e}, CNRS,
24~Rue Lhomond, 75005~Paris, France}
\address{Laboratoire Kastler Brossel, ENS-PSL, CNRS, Sorbonne Universit\'{e},
Coll\`{e}ge de France, 24 Rue Lhomond, 75005~Paris, France}
\author{F. S. Cataliotti}
\address{European Laboratory for Non-Linear Spectroscopy (LENS),
University of Florence, via N. Carrara 1, 50019 Sesto F.no (FI), Italy}
\address{QSTAR and CNR-INO, Largo Enrico Fermi 2, 50125 Firenze, Italy}
\address{Department of Physics and Astronomy, University of Florence, via G. Sansone 1, 50019 Sesto Fiorentino, Italy}
\author{A. Smerzi}
\address{European Laboratory for Non-Linear Spectroscopy (LENS),
University of Florence, via N. Carrara 1, 50019 Sesto F.no (FI), Italy}
\address{QSTAR and CNR-INO, Largo Enrico Fermi 2, 50125 Firenze, Italy}
\date{\today}
\maketitle

\textit{Introduction.}---The dynamical quantum Zeno effect describes an evolution induced by measurement back-action, forcing the system to follow a sequence of projections \cite{Neumann,Misra,Facchi,Smerzi}. This technique provides a robust method for quantum control of populations, as has been demonstrated experimentally, e.g., for the static quantum Zeno effect with trapped ions \cite{Itano} and atoms \cite{Streed}, as well as for dynamical evolutions with atoms \cite{Schafer,HarocheQZ,Barontini}. Interestingly, an evolution driven by quantum back-action may also induce a nontrivial change of the quantum mechanical phase, despite the phase-insensitive nature of projective measurements. Indeed, the evolution along a closed trajectory gives rise to a geometric phase factor, on top of a possible dynamical phase that depends on the realization of the trajectory. A closer inspection reveals that the geometric phase is imprinted by the final projection in the sequence, whereas all previous projections effectively freeze the evolution of the geometric, but not of the dynamical phase.

Geometric phases in quantum physics were first systematically studied for adiabatic evolutions by Berry \cite{Berry} and later generalized to arbitrary periodic evolutions by Aharonov and Anandan \cite{AA87}. They are now an integral part of all fields of quantum physics \cite{Wilczek} and have been observed in several experiments, including nuclear magnetic resonance \cite{Suter}, molecular systems \cite{Mead}, graphene \cite{Zhang}, solid-state qubits \cite{SSQ}, and cold atoms \cite{Webb,Bloch}. Controlling and manipulating geometric phase factors provides a robust alternative to engineering quantum states by purely dynamical evolutions. As such, these techniques have potential applications in the field of quantum information, in particular, quantum simulations \cite{ReviewSimulations} and quantum computations \cite{Leibfried}. Geometric phases that arise in a sequence of projections have first been discussed in the pioneering work of Pancharatnam \cite{Pancharatnam}, and were subject of several subsequent studies \cite{BerryKlein,Hariharan,Bhandari,Pati,FacchiPLA}.

In this work, we experimentally generate and measure geometric phase factors in a Bose-Einstein condensate using a combination of free evolution and the dynamical quantum Zeno effect. We provide the theoretical framework for their interpretation showing that the purely geometric phase generated by a sequence of closely spaced projections on a closed trajectory can be understood by an equivalent representation in two steps. In a first step, the system evolution along the trajectory is interrupted by frequent projections onto the initial state, which effectively freezes the dynamics. In a second step, the trajectory is retraced in the opposite direction, undisturbed by measurements. The geometric phase is entirely acquired as Aharonov-Anandan phase in the second step. Since both steps require the same amount of time, their dynamical phases cancel each other. This leads to a robust method to isolate geometric phase factors from dynamical contributions, even if the Hamiltonian cannot be controlled.

\textit{Dynamical quantum Zeno effect and geometric phases.}---
Consider a family of states $\lbrace|\Psi_k\rangle\rbrace$ each obtained from a small rotation $U(\delta t)$ of the preceding one, i.e., $|\Psi_k\rangle=U(\delta t)|\Psi_{k-1}\rangle$. If the first state $|\Psi_0\rangle$ and the last state $|\Psi_N\rangle=U(\delta t)^N|\Psi_0\rangle=e^{i\phi_0}|\Psi_0\rangle$ are identical except for the phase factor $\phi_0$ then the states $\lbrace|\Psi_k\rangle\rbrace$ form a closed trajectory. Let us now consider a dynamics induced by a series of successive projections onto the states $\lbrace|\Psi_k\rangle\rbrace$, starting from the initial state $|\Psi_0\rangle$. The final state $|\Psi_{\rm f}\rangle$ at the end of the sequence of projections is given by
\begin{equation}\label{eq:projective}
|\Psi_{\rm f}\rangle = |\Psi_N\rangle\langle \Psi_N|\Psi_{N-1}\rangle\cdots\langle\Psi_2|\Psi_1\rangle\langle\Psi_1|\Psi_0\rangle.
\end{equation}
Notice that the above expression can be written as 
\begin{align}\label{eq:projective1}
|\Psi_{\rm f}\rangle = e^{i\phi_0}|\Psi_0\rangle\left(\langle\Psi_0|U^{\dagger}(\delta t)|\Psi_{0}\rangle\right)^N=e^{i\phi}|\Psi_0\rangle,
\end{align}
where the phase $\phi = \phi_0+\phi_{p}$ contains both the phase $\phi_0$ of the free evolution from $|\Psi_0\rangle$ to $|\Psi_N\rangle$ and an additional phase factor from the projections $e^{i\phi_{p}}=\Pi_{k=1}^N\langle\Psi_k|\Psi_{k-1}\rangle=\left(\langle\Psi_0|U^{\dagger}(\delta t)|\Psi_{0}\rangle\right)^N$. This shows that the evolution~(\ref{eq:projective}) can be equivalently decomposed into (i) the \textit{Zeno evolution}: $N$ evolution steps backward in time of duration $\delta t$, each one followed by a projection onto the initial state and finally (ii) the \textit{measurement-free evolution}: one long forward evolution of duration $T=N \delta t$.

\begin{figure}[tb]
\centering
\includegraphics[width=.47\textwidth]{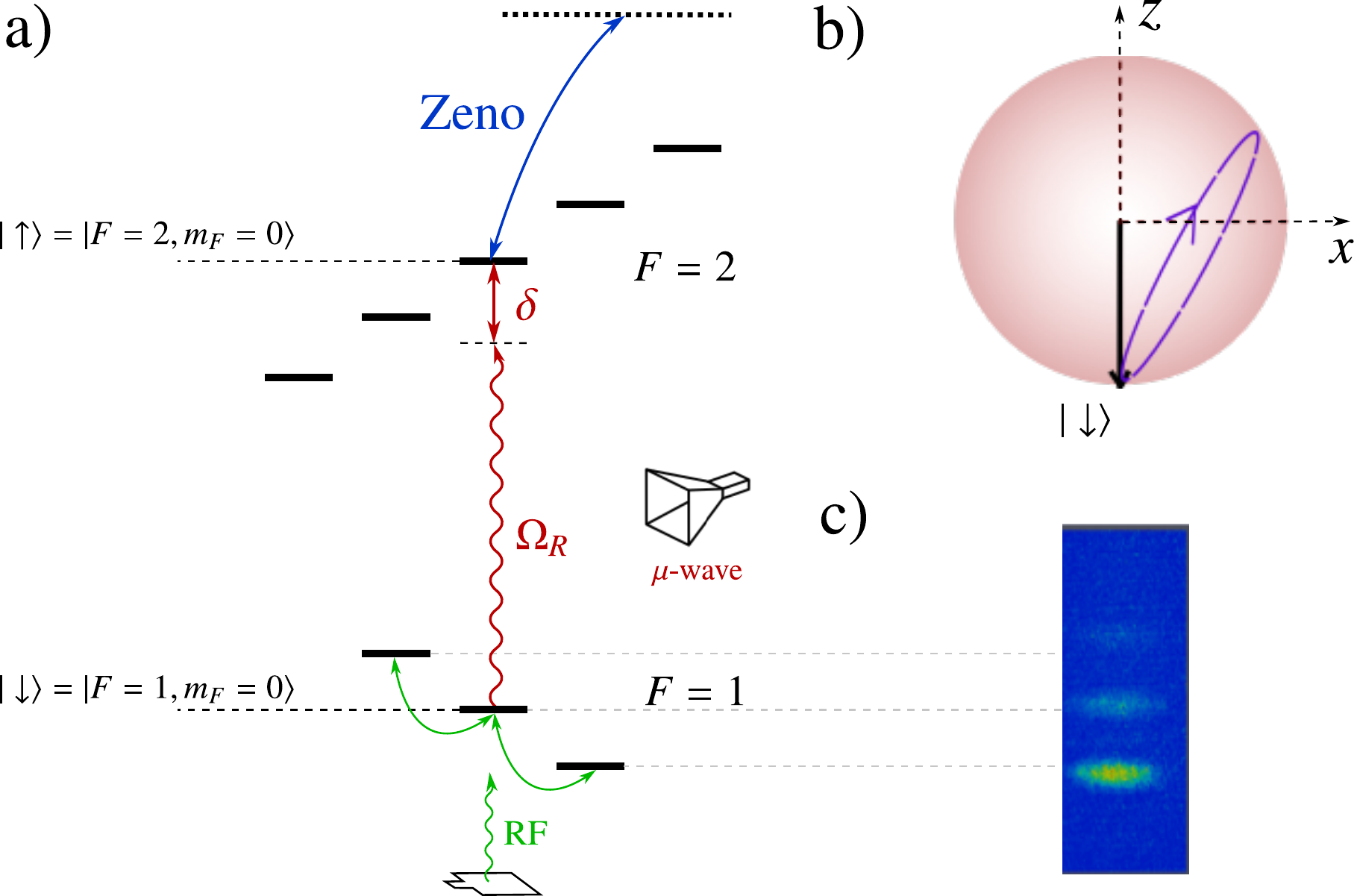}
\caption{a) Schematic representation of the level structure of $^{87}$Rb in the presence of a weak homogeneous magnetic field and the fields used to control the quantum state. A resonant radiofrequency (RF) field couples magnetic sublevels within the same hyperfine state (green lines), a quasi-resonant microwave field couples different hyperfine levels (red line), and an optical field couples realizes the quantum Zeno dynamics by coupling to an excited state (blue line). b) Pictorial representation of the trajectory on the Bloch sphere achieved by the evolution with the Hamiltonian $H_{\mathbf{n}}$. c) Absorption image of the $F=1$ manifold after expansion in a magnetic field gradient. The separation between clouds is $\approx 200 \mu$m.}
\label{fig:1}
\end{figure}

Let us analyze the two steps of the process~(\ref{eq:projective}) separately, focusing first on the \textit{measurement-free evolution} [step (ii)]. We denote by $|\downarrow\rangle$ and $|\uparrow\rangle$ the eigenstates of the Pauli matrix $\boldsymbol{\sigma_z}$. The system evolution is generated by the Hamiltonian $H_{\mathbf{n}}=\omega\frac{\mathbf{n}\cdot\boldsymbol{\sigma}}{2}+\epsilon\frac{\mathbb{I}}{2}$, where $\mathbf{n}=(\sin\theta,0,\cos\theta)$ is a unit vector and we include an arbitrary energy offset $\epsilon$ whose significance will be specified below. After a period of $T=2\pi/\omega$, the state $|\Psi_0\rangle=|\downarrow\rangle$ evolves with $U(t)=e^{-iH_{\mathbf{n}}t}$ to the state $|\Psi(T)\rangle=U(T)|\downarrow\rangle=e^{i\phi_0}|\downarrow\rangle$ and acquires a total phase of $\phi_0=-\pi(1+\frac{\epsilon}{\omega})$, which follows from $U(T)=e^{-i\pi(1+\frac{\epsilon}{\omega})}\mathbb{I}$. Following Aharonov and Anandan \cite{AA87} we determine the dynamical phase associated with this evolution as 
\begin{align}\label{eq:dynphase}
\phi_d=-\int_0^T dt\langle\Psi(t)|H_{\mathbf{n}}|\Psi(t)\rangle=\pi\left(\cos\theta-\frac{\epsilon}{\omega}\right).
\end{align}
The total phase is now decomposed as $\phi_0=\phi_d+\beta$, with the fully geometric Aharonov-Anandan phase
\begin{align}
\beta=\Omega/2,
\label{SolidAngle}
\end{align}
where $\Omega=2\pi(1-\cos\theta)$ is the solid angle of the trajectory subtended at the origin of the Bloch sphere. The phase $\beta$ is therefore independent of the energy offset $\epsilon$, while the dynamical and the total phases are not.

Next, we analyze the phase $\phi_{p}$ that arises from the sequence of projections: the \textit{Zeno evolution} [step (i)]. We obtain $e^{i\phi_p}=\langle\downarrow|U^{\dagger}(\delta t)|\downarrow\rangle^N=\left(\cos(\pi/N)-i\cos\theta\sin(\pi/N)\right)^Ne^{i\frac{\epsilon}{\omega}\pi}$, which in the limit of $N\rightarrow\infty$ tends to (see, e.g., \cite{BerryKlein})
\begin{align}
\lim_{N\to\infty}\phi_p=-\pi\left(\cos\theta-\frac{\epsilon}{\omega}\right)=-\phi_d.
\end{align}
This corresponds to the negative dynamical phase of the \textit{measurement-free evolution}~(\ref{eq:dynphase}). The negative sign is due to the appearance of adjoint evolution operators $U^{\dagger}(\delta t)$ that propagate backwards in time. 

Combining the two results, the purely \textit{projective evolution}~(\ref{eq:projective}) indeed imprints only the geometric phase $\phi=\beta$, as pointed out in Refs.~\cite{BerryKlein,FacchiPLA}. Our analysis reveals that the origin of the geometric phase can be traced back to the Aharonov-Anandan phase that is generated in the final evolution [step (ii)]. Any dynamical phase that is created in this step is canceled by the phase accumulated during the projections [step (i)], as both steps require the same time $T$.

\textit{Experimental realization.}---We produce a Bose-Einstein condensate of $\approx 10^5$ $^{87}$Rb atoms in $|F=2,m_F=2\rangle$ in a magnetic microtrap with frequencies $f_{axial}=76\rm\,Hz$ and $ f_{radial}=975\rm\,Hz$ realized with an atom chip~\cite{Petrovic}. After switching off the magnetic trap, we expand for $1\,\rm ms$ so that we can neglect the effect of atomic interactions. Then we lift the magnetic degeneracy of the hyperfine levels by applying a homogeneous and constant magnetic field of $6.179 \rm\,G$ to the condensed atoms. Thanks to the opposite sign of the Land\'e factors in the two hyperfine ground levels, this also effectively isolates different closed 2-level systems in the $|F=1\rangle\rightarrow|F=2\rangle$ microwave transition. 

With a frequency modulated radiofrequency (RF) pulse designed with an optimal control strategy \cite{Lovecchio} we transfer all the atoms into the $|\uparrow\rangle=|F=2;m_F=0\rangle$ state. Subsequently a microwave $\pi$-pulse at $\omega_0=6.834\,703\,\rm GHz$ is used to transfer all the atoms in $|\downarrow\rangle=|F=1;m_F=0\rangle$. This is the initial state $|\Psi_0\rangle$ for our experiment. 
By applying an RF $\pi/2$-pulse at $4.323\,\rm{MHz}$, resonantly coupling neighboring $m_F$ states, we produce a superposition $|\Psi\rangle=(|\downarrow\rangle+|\Psi_r\rangle)/\sqrt{2}$, where $|\Psi_r\rangle=(|1;-1\rangle+|1;1\rangle)/\sqrt{2}$ is used as a local oscillator to provide a phase reference (see Ref.~\cite{Gross} for atomic local oscillator techniques based on strongly populated coherent states).

We then drive the hyperfine transition $|1;0\rangle\rightarrow|2;0\rangle$ with a microwave field of frequency $\omega_L$.
By defining the resonant Rabi frequency produced by the microwave as $\Omega_R$ and the detuning $\delta=\omega_0-\omega_L$, the atomic evolution is described in the rotating wave approximation by the Hamiltonian $H_{\mathbf{n}}$, where we identify $\omega=\sqrt{\Omega_R^2+\delta^2}$ and $\cos\theta=\delta/\omega$ and $\sin\theta=\Omega_R/\omega$.  
We realize the projective measurements $P=|\Psi_0\rangle\langle\Psi_0|$ by illuminating the atoms with a light pulse of $1.5\rm\,\mu s$ duration, resonant with the $|F=2\rangle\rightarrow|F=3\rangle$ component of the Rubidium $D2$ line. This light pulse would lead to the loss of the atoms that were in the $|F=2\rangle$ state, effectively implementing a projection onto the subspace orthogonal to $|\Psi_0\rangle$. The frequent repetition (pulses every $\tau=2\rm\,\mu s\ll 1/\omega$) of such pulses effectively prevents atoms from entering this subspace hence realizing the quantum Zeno dynamics.

After a period $T$, at the end of the evolution, the phase is measured by overlapping with the local oscillator using a second RF $\pi/2$-pulse. 
In essence, this closes a 3-level Ramsey-like atomic interferometer. In the absence of driving, the interferometer output exhibits fringes in the atomic population of the three levels with a periodicity dictated by the homogeneous magnetic field. Moreover, a preliminary test shows that the Zeno regime was attained (see Appendix~\ref{app} for details).

To record the number of atoms in each of the internal states, we use a Stern-–Gerlach method. To separate the different $m_F$ states, after 1\,ms of free expansion in addition to the homogeneous bias field, we apply a magnetic field gradient of $4$\,G/cm along the quantization axis for $10$\,ms.  After further $13$\,ms of expansion, a standard absorption imaging sequence is executed. To ensure a strong absorption process, and thus a large signal-to-noise ratio, the probe beam is
resonant with the transition $|F = 2\rangle \rightarrow |F = 3\rangle$. We take an image of atoms in $|F=2\rangle$, then repump atoms from $|F=1\rangle$ to $|F=2\rangle$ to take a second image $25\rm\,\mu s$ later.

With the atoms always initialized in the state $|\downarrow\rangle = |\Psi_0\rangle$, the experimental protocol consists in recording the interferometric fringes for the following four cases:
\begin{enumerate}
    \item \textit{Reference evolution 1}: Evolution of the system in absence of coupling between $|\downarrow\rangle$ and $|\uparrow\rangle$ accumulating a phase shift $\phi_1$ on $|\downarrow\rangle$.
    \item \textit{Reference evolution 2}: Evolution of the system in absence of coupling between $|\downarrow\rangle$ and $|\uparrow\rangle$ while performing projective measurements onto $|\downarrow\rangle$ accumulating a phase shift $\phi_2$ on $|\downarrow\rangle$.
    \item \textit{Measurement-free evolution}: Evolution of the system in the presence of coupling $|\downarrow\rangle$ and $|\uparrow\rangle$ with $H_{\mathbf{n}}$ accumulating a phase shift $\phi_3$ on $|\downarrow\rangle$.
    \item \textit{Zeno evolution}: Evolution of the system in the presence of coupling $|\downarrow\rangle$ and $|\uparrow\rangle$ with $H_{\mathbf{n}}$ while performing projective measurements onto $|\downarrow\rangle$ accumulating a phase shift $\phi_4$ on $|\downarrow\rangle$.
\end{enumerate}
All experiments are repeated 5 times for statistical averaging. 

Cases 1 and 2 are theoretically and experimentally identical, thus proving that there is indeed no disturbance on the local oscillator coming from the projective measurements. All the experimentally measured phases in the following are extracted by comparing to either of these references (see Appendix~\ref{app} for further details). In the theoretical description, the phase reference is taken into account via the arbitrary energy offset $\epsilon$.

\begin{figure}[tb]
\centering
\includegraphics[width=.46\textwidth]{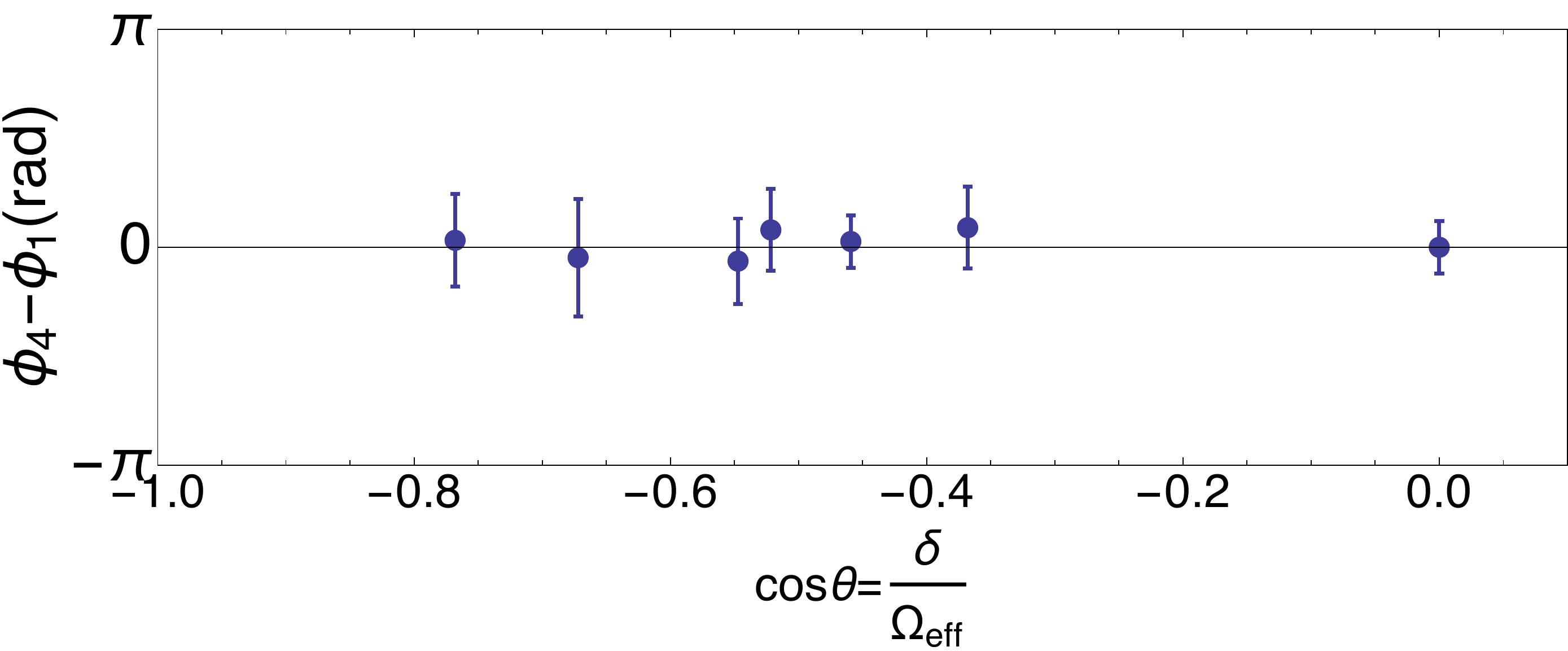}
\caption{Difference between phase shifts recorded for the \textit{Zeno evolution} (case 4) and the purely dynamical phase that is obtained by the \textit{reference evolution} without driving and measurements (case 1).}
\label{fig:2}
\end{figure}

In Fig.~\ref{fig:2} we show the experimentally recorded phases for the \textit{Zeno evolution} [step (i)] for different detunings $\delta$,  hence rotating $\mathbf{n}$. Here we show the forward evolution driven by $H_{\mathbf{n}}$ that generates the phase $-\phi_p=\phi_d$. The projective part [step (i)] in Eq.~(\ref{eq:projective}) describes an evolution along the same trajectory in opposite orientation which corresponds to the dynamics generated by $H_{-\mathbf{n}}$. The measured phases $-\phi_p$ are compared to the phases obtained from waiting for the same time $T$ without any external driving or measurements (\textit{reference evolution 1}), which imprints the dynamical phase $-(\epsilon-\delta)T/2=\pi(\delta-\epsilon)/\omega=\phi_d$, as defined in Eq.~(\ref{eq:dynphase}), with $\epsilon$ given by the RF frequency. The coinciding phase shifts for these two evolutions confirm that the projections [step (i)] freeze the system in the initial state, which inhibits the accumulation of geometric phase factors, while a dynamical phase is still acquired due to the passing of time.

\begin{figure}[tb]
\centering
\includegraphics[width=.47\textwidth]{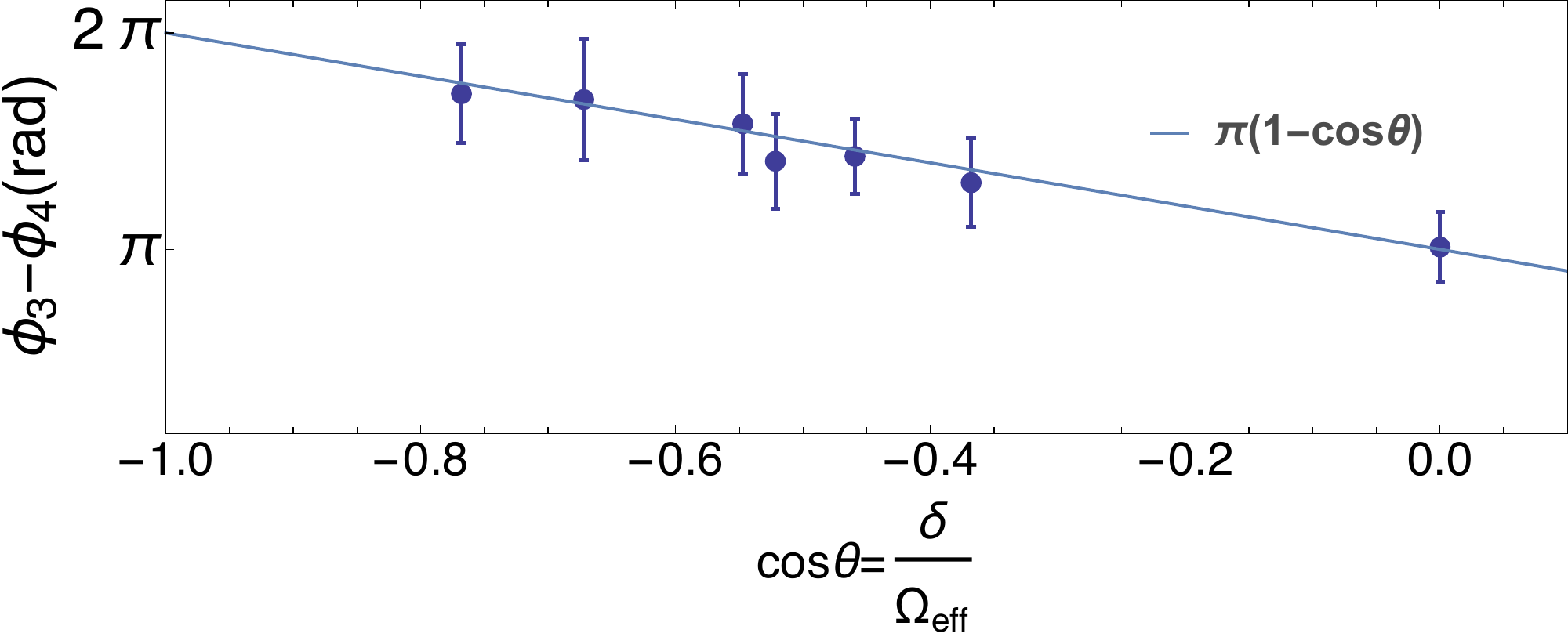}
\caption{Difference between phase shifts of the \textit{measurement-free evolution} in the presence of coupling without Zeno measurements (case 3) and the phase shift coming from the \textit{Zeno evolution} in the presence of coupling with Zeno measurements (case 4). The phase difference $\phi_0-\phi_d=\beta$ is the purely geometric Aharonov-Anandan phase, as predicted in Eq.~(\ref{SolidAngle}), that only depends on the trajectory and is independent of the dynamical process.}
\label{fig:3}
\end{figure}

Figure~\ref{fig:3} shows the phase shifts generated by the closed trajectory of projections, described in Eq.~(\ref{eq:projective}). This corresponds to the combination of both steps (i) and (ii). We show the result of the subtraction of the phase shift acquired during the \textit{Zeno evolution} ($\phi_4$) from the phase shift acquired during the \textit{measurement-free evolution} ($\phi_3$) for various detunings $\delta$ of the microwave coupling.
The result, as predicted in Eq.~(\ref{SolidAngle}), is the purely geometric Aharonov-Anandan phase $\beta=\pi\left(1-\cos\theta\right)$ shown as a continuous line in the figure. 

\begin{figure}[tb]
\centering
\includegraphics[width=.47\textwidth]{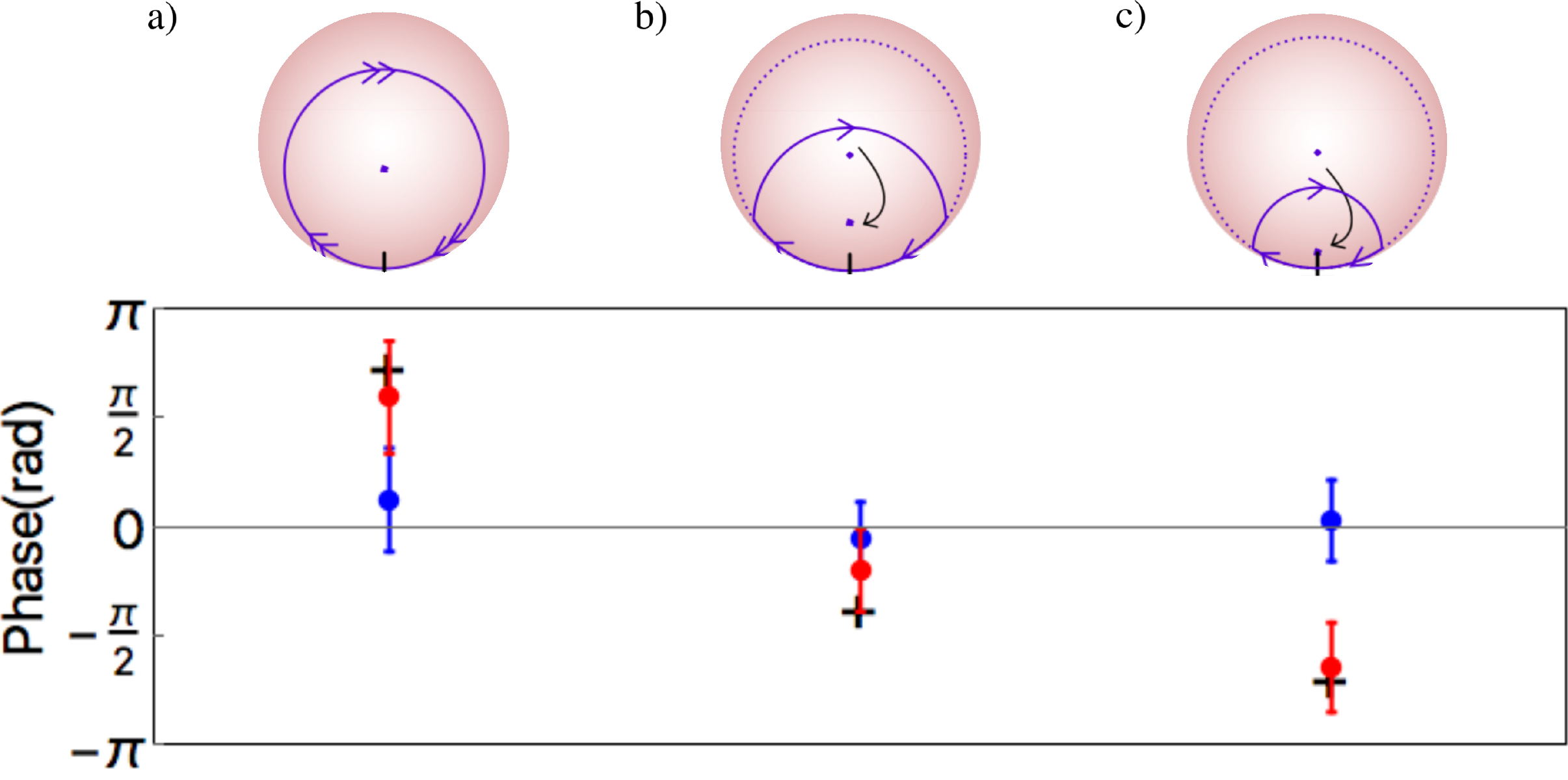}
\caption{Phase shifts for the different trajectories represented on the Bloch sphere at the top of the figure. The three cases a), b), and c) are explained in the text. The blue dots show the differences ($\phi_4-\phi_1$) between phase shifts recorded for the \textit{Zeno evolution} (case 4) and the purely dynamical phase that is obtained by the \textit{reference evolution 1} without driving and measurements (case 1). The difference is expected to be compatible with zero (gray line). The red dots show the differences ($\phi_3-\phi_4$) between phase shifts for the \textit{measurement-free evolution} in the presence of coupling without Zeno measurements (case 3) and the phase shift coming from the \textit{Zeno evolution} in the presence of coupling with Zeno measurements (case 4). This difference realizes the purely projective evolution~(\ref{eq:projective}). The expected geometrical phase factor is predicted by Eq.~(\ref{SolidAngle}) which is shown as black dots.}
\label{fig:4}
\end{figure}

To demonstrate that the measured phase $\beta$ is indeed geometric and only depends on the properties of the trajectory via Eq.~(\ref{SolidAngle}), we perform a series of additional measurements with different trajectories. As shown in Fig.~\ref{fig:4}, we performed three sets of experiments: 
\begin{enumerate}[label=\alph*)]
    \item Retracing a circular trajectory twice.
    \item Evolving along a larger intersection of two spherical caps by abruptly changing $\mathbf{n}$ during the evolution.
    \item Evolving along a smaller intersection of two spherical caps by abruptly changing $\mathbf{n}$ during the evolution.
\end{enumerate}
For case a), we expect to find twice the solid angle of the single trajectory. The detuning was chosen as $\delta=16\,\rm kHz$, with a resonant Rabi frequency $\Omega_R=40.4\,\rm kHz$ leading to an angle of $\theta=1.19\,\rm rad$. The solid angles of intersecting spherical caps b) and c) can be determined from elementary geometric considerations. In the experiment b) was realized with a fixed detuning of $\delta=24\,\rm kHz$ by suddenly changing the resonant Rabi frequency from $\Omega_R=36.7\,\rm kHz$ to $\Omega_R=26.4\,\rm kHz$ and back so that the system still evolves on a closed-loop. Case c) describes a fixed detuning of $\delta=16\,\rm kHz$ and a sudden transition of the Rabi frequency from $\Omega_R=36.7\,\rm kHz$ to $\Omega_R=19.9\,\rm kHz$ and back, again ensuring that the system evolves on a closed-loop.

The corresponding trajectories on the Bloch sphere are pictorially represented in Fig.~\ref{fig:4}. For all three cases we performed a \textit{reference evolution 1} obtaining the phase $\phi_1$, a \textit{measurement-free evolution} obtaining the phase $\phi_3$ and a \textit{Zeno evolution} obtaining the phase $\phi_4$. The difference of the measured phases is in satisfactory agreement with the theoretical prediction of Eq.~(\ref{SolidAngle}), which confirms their geometric nature.

\textit{Conclusions.}---Quantum evolution typically leads to phase shifts with mixed geometric and dynamical origins. It is necessary to eliminate the dynamical phase from the total phase to identify geometric phases. In this manuscript, we have experimentally generated and measured geometric phases in a system of cold trapped atoms by realizing a sequence of closely spaced projectors along with a closed trajectory. This evolution consists of a combination of dynamical quantum Zeno effect and free evolution. By analyzing the phase shifts generated in each step separately, we demonstrate -- both theoretically and experimentally -- that the geometric phase is imprinted in the free evolution as an Aharonov-Anandan phase. On the other hand, by frequent projective measurements, it is possible to effectively inhibit the formation of the geometric phase without disturbing the dynamical phase. The quantum Zeno effect thus represents a robust method to separate dynamical and geometric contributions of the quantum phase.

Our method gives us two possibilities to reveal the geometric Aharonov-Anandan phase generated by the free evolution. Either we can compare the evolution to the free evolution that is obtained by ``switching off'' the driving Hamiltonian, or we can disable the formation of a geometric phase by realizing quantum Zeno dynamics through frequently repeated measurements. Both methods enable us to identify and remove the contribution of the dynamical phase from the total phase. The latter possibility has particularly interesting applications since it does not require the ability to control the Hamiltonian. 

The presented method enables the measurement and identification of geometric phases in systems where it is typically concealed by dynamical phases that cannot be readily determined or eliminated. As such, it may lead to exciting novel approaches for the identification of topological phases of matter \cite{Xiao,Qi} in synthetic systems \cite{Bloch,Cooper} and potential applications in the context of quantum phase transitions \cite{Zhu} or quantum computations \cite{Nayak,Vedral,Bharath}.

\textit{Acknowledgments.---}We thank S. Pascazio for useful discussions and M. Inguscio for continuous support. The European Commission has supported this work through the QuantERA ERA-NET Cofund in Quantum Technologies project ``Q-Clocks''. M.G. acknowledges funding by the Alexander von Humboldt Foundation and the LabEx ENS-ICFP:ANR-10-LABX-0010/ANR-10-IDEX-0001-02 PSL*.

\appendix

\section{Experimental details on the phase measurements}
\label{app}
A homogeneous magnetic field of absolute value $B$ lifts the degeneracy of the three magnetic sublevels corresponding to $m_F=-1,0,1$ of the hyperfine ground states $|F=1\rangle$ of $^{87}$Rb atoms. The frequency difference $\Delta_{\pm}=\nu_{\pm 1}-\nu_0$  between the $m_F=\pm 1$ and the $m_F=0$ sublevels depends on $B$ as:
\begin{equation}
\begin{aligned}
    \Delta_{\pm}= \mp g_I\mu_B B -\frac{\nu_{\downarrow \uparrow}}{2}\sqrt{1\pm \frac{(g_J - g_I)\mu_B B}{\nu_{\downarrow \uparrow}}+\left(\frac{(g_J - g_I)\mu_B B}{\nu_{\downarrow \uparrow}}\right)^2},\notag
\end{aligned}
\end{equation}
where $g_I$ and $g_J$ are Land\'e factors, $\mu_B$ is the Bohr magneton and $\nu_{\downarrow \uparrow}$ is the frequency of the transition between the  $|1,0\rangle=|\downarrow\rangle$ and  $|2,0\rangle=|\uparrow\rangle$ states.

\begin{figure}[tb] 
\centering
\includegraphics[width=.45\textwidth]{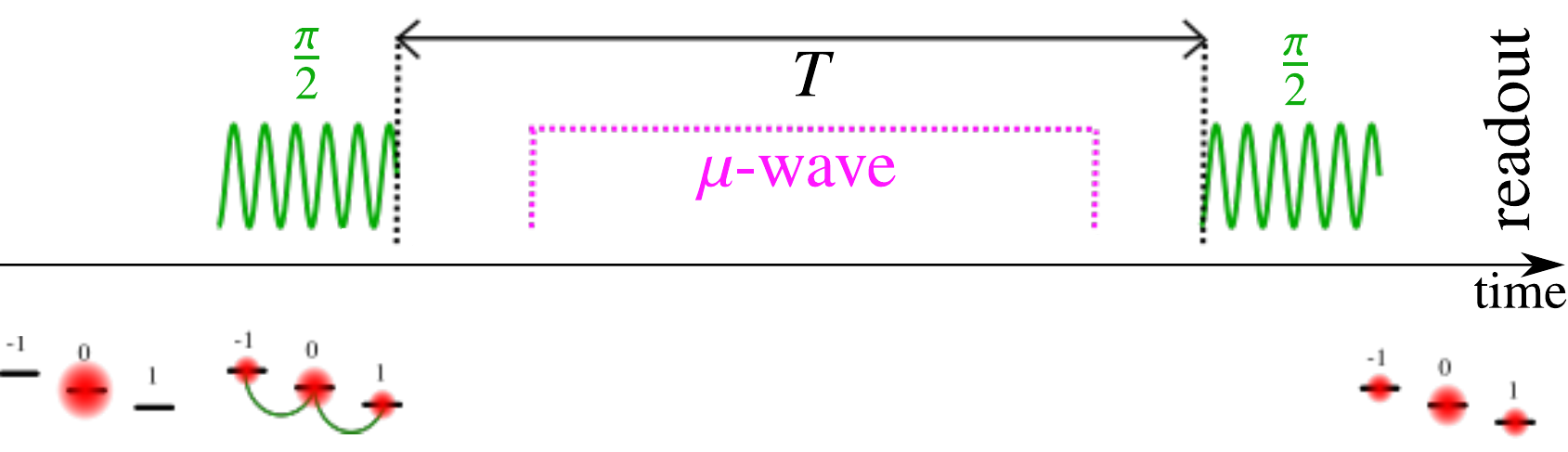}
\caption{
Scheme of the three-level atomic interferometer. A first $\pi/2$ RF pulse (composed of 27 cycles and a total duration of $6.245\,\mu$s) transfers half of the atomic population from the $|1,0\rangle$ state to the $|1,\pm 1\rangle$ states. These levels are unaffected either by the microwave coupling or by the projective measurements. Then the system evolves for a time $T$ in one of the 4 cases discussed in the main text, which may involve driving with a microwave field. Finally, a second $\pi/2$ RF pulse closes the interferometer, and the populations are readout.
}
\label{levels}
\end{figure}

In the experiment, the initial state is the central ${|F = 1, m_F = 0\rangle}=|0,1\rangle$ state. Coherent transfer of excitations between magnetic sublevels of the same hyperfine state is realized by the
application of a near-resonant RF pulse with frequency $|\Delta_{\pm}|\approx 4.323\,\rm{MHz}$. We realize a Ramsey interferometer \cite{Ramsey} by applying two RF $\pi/2$ pulses, separated by a controllable time delay $T$ (see Fig.~\ref{levels}). The second pulse maps the relative
phases accumulated between different states during the delay into a population distribution at
the output of the interferometer. At the end of the sequence, we measure the populations of the three magnetic levels.

\begin{figure}[tb]
\centering
\includegraphics[width=.49\textwidth]{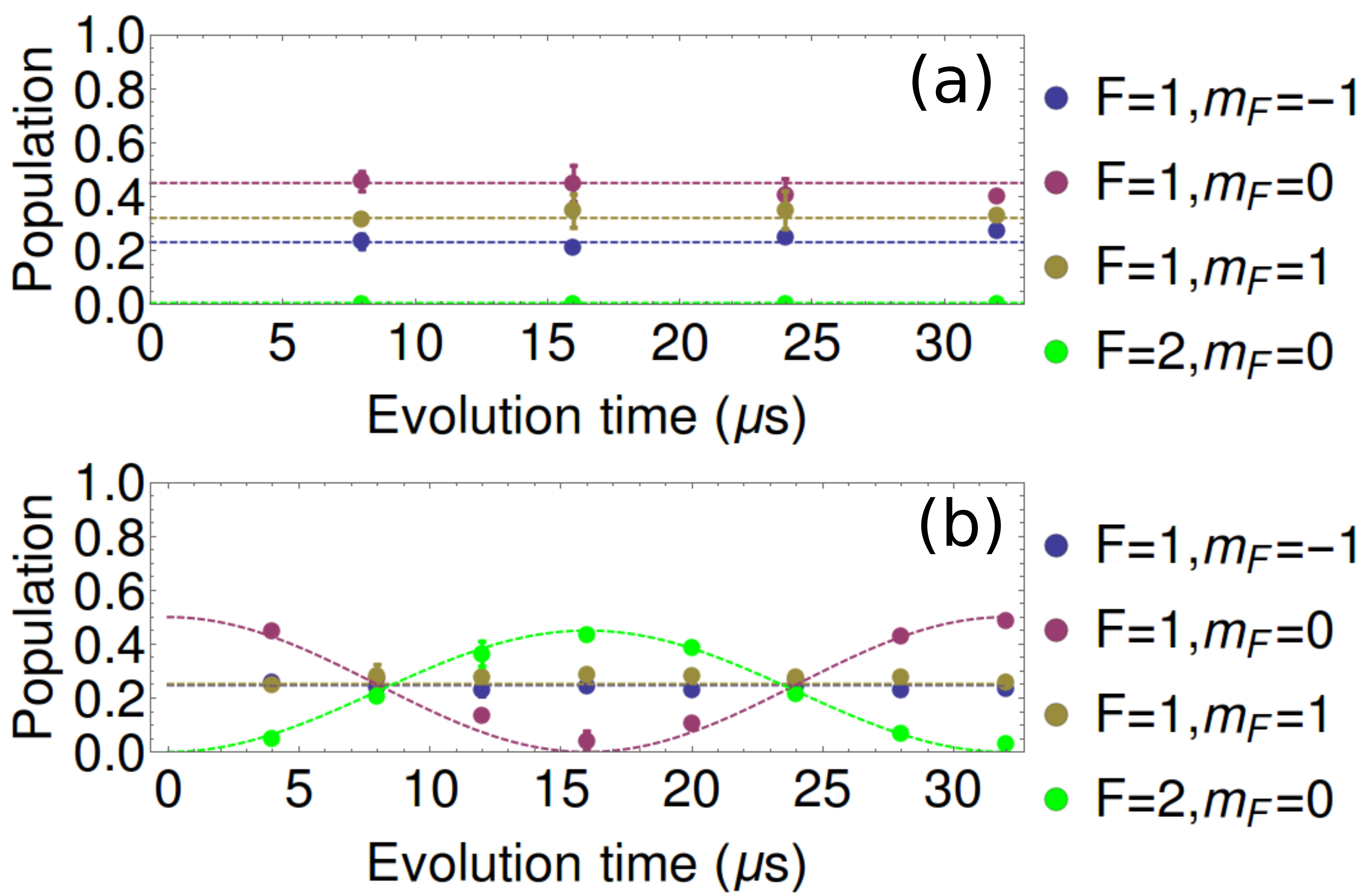}
\caption{(a) Evolution of the atomic population in the presence of Zeno projective measurements. (b) Reference evolution without Zeno measurements.}
\label{Zenotest}
\end{figure}

We first confirm the effectiveness of the Zeno pulses ($1.5\rm\,\mu s$ of light resonant with $F=2 \rightarrow F'=3$ transition, see main text) for the experiment. We divide the atoms with the RF-pulse in the three levels of $|F=1\rangle$ (note that the calibration of the RF pulses was not yet perfected in this preliminary test, causing the initial population ratio between $m_F=\{-1,0,1\}$ to slightly deviate from $\{25\%,50\%,25\%\}$). We then measure the population distribution in the presence of the coupling between $|F=1,m_F=0\rangle$ and $|F=2,m_F=0\rangle$. We measure the population at each level while driving the microwave transition around one full Rabi cycle. Afterwards, we repeat the same experiment while, at the same time, applying Zeno pulses every $2\rm\,\mu s$. We see from Fig.~\ref{Zenotest} that the Zeno inhibited the atoms in $|F=1, m_f=0\rangle$ from going to $|F=2,m_F=0\rangle$.

In the main experimental sequence, the first RF $\pi/2$-pulse produces the superposition $|\Psi\rangle=({|\downarrow\rangle}+|\Psi_r\rangle)/\sqrt{2}$, where $|\Psi_r\rangle=(|1;-1\rangle+|1;1\rangle)/\sqrt{2}$. The reference state $|\Psi_r\rangle$ is unaffected either by the microwave coupling or by the projective measurements and therefore provides a phase reference. 

\begin{figure}[tb]
\centering
\includegraphics[width=.49\textwidth]{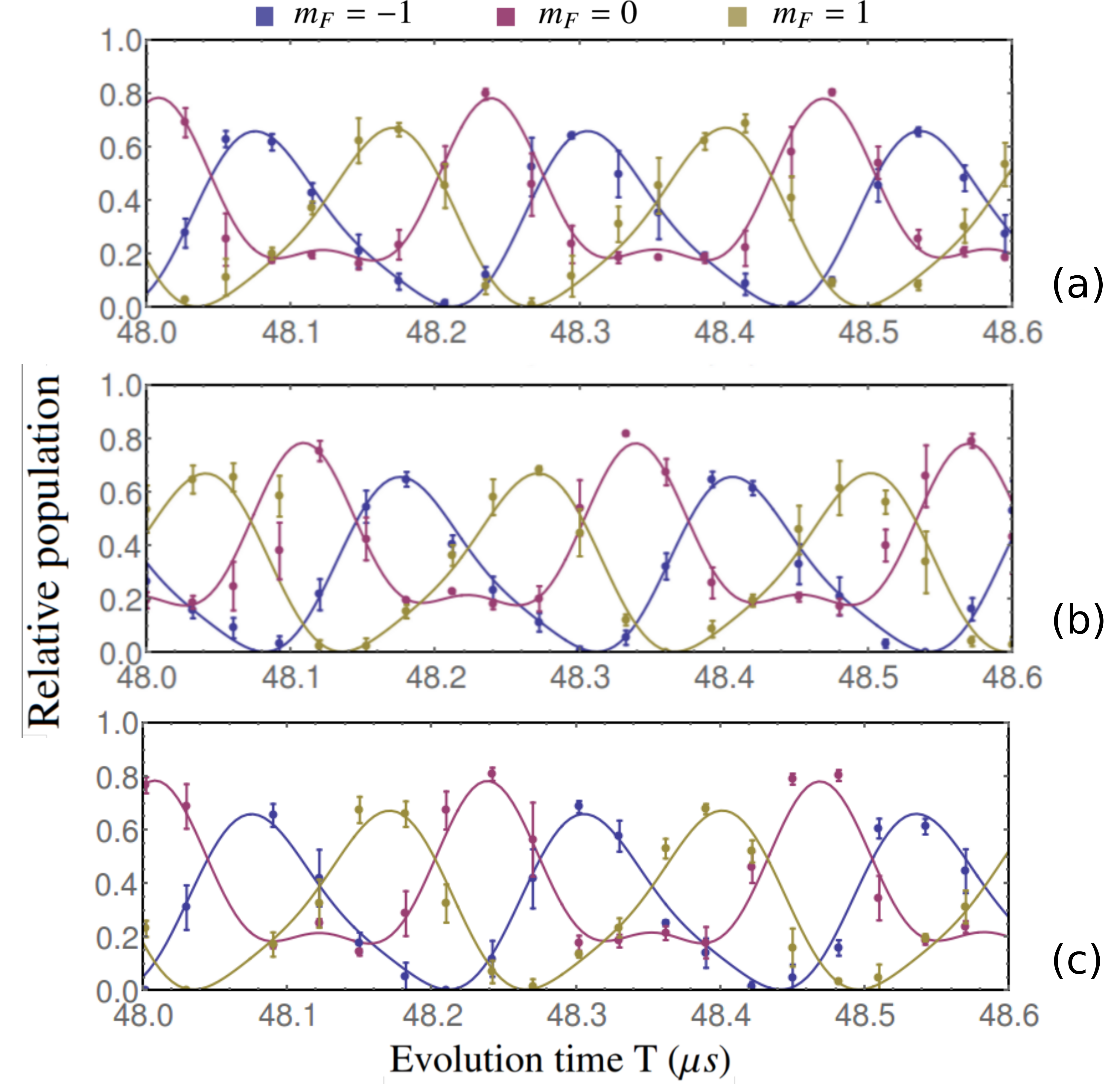}
\caption{Relative populations of the three magnetic sublevels of $|F=1\rangle$ as a function of the evolution time $T$. The experimental dots are the average of 5 realizations with error bars representing the standard deviation. The continuous line is a fit to the data obtained by computing the system dynamics with $B$ as a free parameter in a), without the presence of projective measurements. Next, the magnetic field is fixed to the value found in a) while the free parameter is the phase $\phi$ of level $|1,0\rangle=|\downarrow\rangle$, without the presence of projective measurements in panel b) and with the presence of projective measurements in panel c).}
\label{Interf}
\end{figure}

To record the number of atoms in each of the internal states, we apply a Stern–-Gerlach method~\cite{SG}. After the interferometric procedure, we let the atoms fall in the presence of an inhomogeneous magnetic field applied along the quantization axis. This causes the different $m_F$ states to spatially separate. In the end, a standard absorption imaging sequence is executed~\cite{Ketterle}. Since we are working with identical, non-interacting atoms, the relative atomic population of the three sublevels is equivalent to the probability for each atom to be found in each of the three sublevels.

In Fig.~\ref{Interf}, we report the relative population of the three levels as a function of the evolution time $T$. In Fig.~\ref{Interf}~a) this is shown for one instance of the \textit{reference evolution} (cf. main text). We compute the system dynamics and fit the measured fringes using the magnetic field $B$ as a free parameter. By measuring the magnetic field directly with the atomic interferometer, we can eliminate all systematic effects coming from day to day variations of the system environment. Indeed the absolute value of the magnetic field is found to be fluctuating in the range $6.180 - 6.184\,\rm{G}$, accounting for good stability of the experiment.

In Fig.~\ref{Interf}~b), we present an example of the \textit{measurement-free evolution} (cf. main text) in the case of full rotation in the $yz$ plane. As expected, the fringes are shifted by a half period indicating a phase difference of $\pi$ comparing to the \textit{reference evolution}. 

In Fig.~\ref{Interf}~c) we display an example of the \textit{Zeno evolution} (cf. main text), the absence of contrast deterioration comparing to the \textit{reference evolution} shows that atomic losses induced by the repeated projective measurements were negligible. 

In order to deduce the phase shift $\phi$ in Fig.~\ref{Interf}~b) and ~c), we again compute the system dynamics this time fixing the value of magnetic field B to that obtained in the \textit{reference evolution} and fit the measured fringes using the additional phase $\phi$ as a free parameter. The fit error on the free parameter is evaluated with the standard deviation of the experimental data around the fitting value $\langle \delta P(\phi)^2 \rangle$ and the slope of the fitted population $P(\phi)$ around the fit parameter:

\begin{equation}
\begin{aligned}
    \langle \delta \phi^2 \rangle &= \left<  \left(\delta P(\phi) \frac{\partial \phi}{\partial P(\phi)}\right)^2 \right> \\
    &\approx \langle \delta P(\phi)^2 \rangle \left(\frac{\Delta \phi}{P(\phi + \Delta \phi) - P(\phi)}\right)^2
\end{aligned}
\end{equation}

\end{document}